# Magnetocaloric Effect in Nanostructured $La_{0.6}Sr_{0.4}Fe_{1-x}Co_xO_3$ (x = 0, 0.2, 0.5, 0.8 and 1)

*Fabiana N Morales Alvarez[1,2], Mariano Quintero[1,2], Joaquín Sacanell[1,2]*
*[1]Departamento de Física de la Materia condensada, Gerencia de Investigación y Aplicaciones, Centro Atómico Constituyentes, CNEA. Av. General Paz 1499, (1650) Villa Maipú, Provincia de Buenos Aires, Argentina.*
*[2]Instituto de Nanociencia y Nanotecnología, Centro Atómico Constituyentes, CNEA-CONICET. Av. General Paz 1499, (1650) Villa Maipú, Provincia de Buenos Aires, Argentina.*
* Corresponding author:

## Abstract

This work presents a systematic study of the magnetocaloric effect in the nanostructured perovskite series $La_{0.6}Sr_{0.4}Fe_{1-x}Co_xO_3$ (x = 0, 0.2, 0.5, 0.8, and 1.0), synthesized by a pore-wetting method using polymeric membranes with pore diameters of 200 nm and 800 nm. All samples were calcined at 1000 °C. Structural characterization was made by X-ray diffraction and confirmed the formation of a single-phase perovskite with distorted rhombohedral symmetry with $R\bar{3}c$ space group, without detectable secondary phases. We observed significant influence of substitution of Fe by Co on the morphology, as the analysis by scanning electron microscopy revealed a clear evolution from smaller to larger particles and from thin to thicker nanotubes, as the Co content increased. Magnetic measurements showed that the cationic substitution enhances ferromagnetic coupling, increasing both the saturation magnetization ($M_S$) and the Curie temperature ($T_C$). The magnetocaloric properties, determined through the Maxwell relations, exhibit a maximum entropy change of 1.13 J $kg^{-1}$ $K^{-1}$ under an applied field of 3 T for the sample with x = 1. These results demonstrate that the combination of Co doping and controlled nanostructuring effectively optimizes the magnetocaloric response.

**Highlights**

- Nanorods cobalt-doped perovskites show tunable magnetocaloric effect
- Controlled cobalt substitution strengthens ferromagnetic order
- Morphology engineering enhances magnetic entropy change
- Second-order magnetic transitions confirmed by Arrott analysis
- Peak entropy changes of 1.13 $Jkg^{-1}K^{-1}$ at 240 K

**Graphical Abstract**

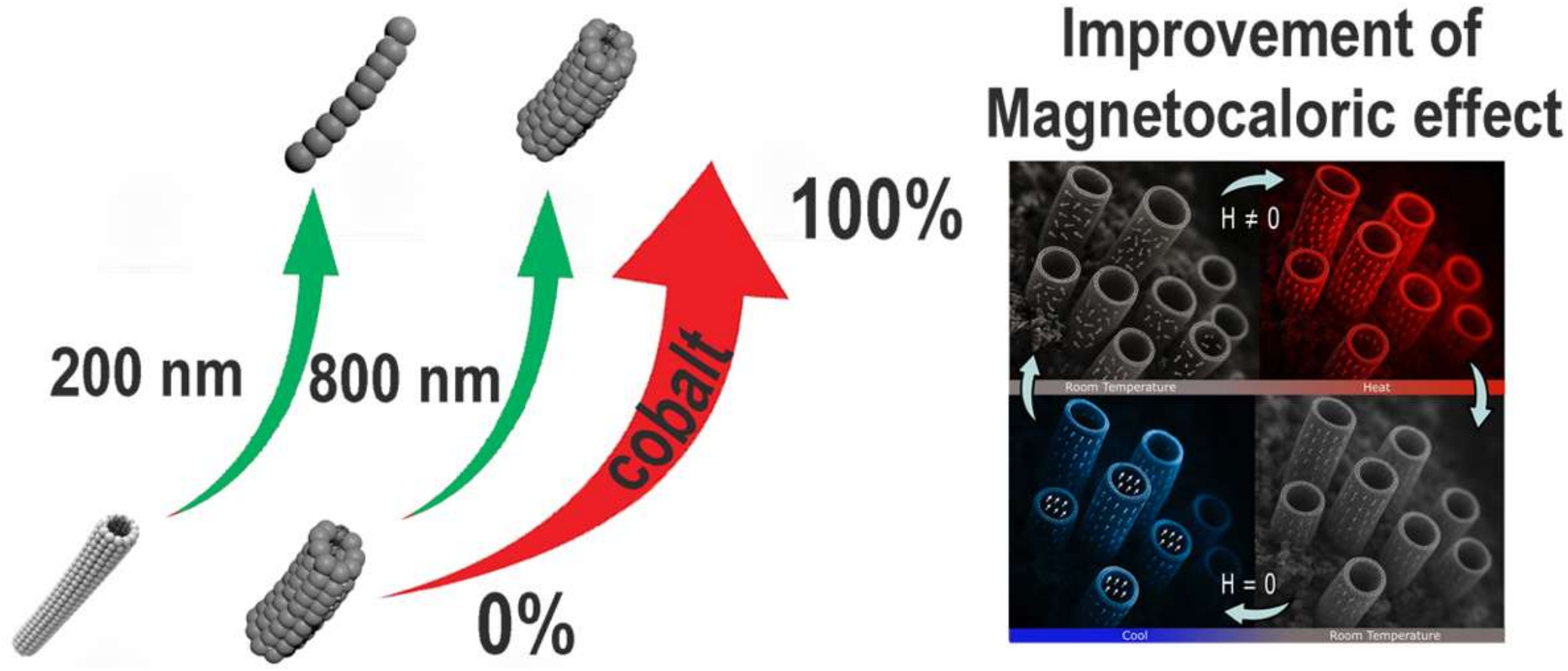

## 1. Introduction

The study of the magnetocaloric effect (MCE) in materials has gained increasing attention over the last decades due to its potential application in magnetic refrigeration technologies[1]. The MCE is defined as the isothermal change in magnetic entropy (−ΔS) that occurs when a magnetic field is applied to a material. [2] This quantity can be derived from the temperature dependence of the magnetization M(H,T) through the Maxwell relation:

$$\left.\frac{\delta S}{\delta H}\right|_T = \left.\frac{\delta M}{\delta T}\right|_H$$

which, in its integral form, can be expressed as:

$$\Delta S(T,H) = \int_0^H \left(\frac{\partial M(H,T)}{\partial T}\right)_H dH'$$

by numerical integration

$$\Delta S(T,H) = \int_0^H \left(\frac{[M(T+\Delta T,H') - M(T,H')]}{\Delta T}\right)_H dH'$$

Here, M(T,H′) represents the magnetization measured at temperature T and field H′, while M(T + ΔT,H′) corresponds to the magnetization at a slightly higher temperature under the same field. From these data, the magnetic entropy change (−ΔS) is calculated[2].

Numerous studies have established a strong correlation between the magnetic properties and the MCE: materials with high saturation magnetization ($M_S$) and low coercive fields ($H_C$)—such as superparamagnetic systems—tend to exhibit efficient magnetization and demagnetization cycles, resulting in higher entropy change ($-\Delta S$) and relative cooling power (RCP) values [2]. Therefore, materials exhibiting significant $M_S$ and low $H_C$ values are of particular interest. Among them, cobalt-based transition-metal oxides stand out due to their rich magnetic and electronic transport behavior, especially within the $ABO_3$-type perovskite framework. In the cobaltite systems $R_{1-x}A_xCoO_3$, spin-state transitions of Co ions and exchange interactions between $Co^{3+}$–$O^{2-}$–$Co^{3+}$ and $Co^{3+}$–$O^{2-}$–$Co^{4+}$ pairs can produce both antiferromagnetic (AFM) and ferromagnetic (FM) phases, leading to inhomogeneous ferromagnetism below the Curie temperature ($T_C$) [3–5].

Transition-metal oxides crystallizing in the $ABO_3$ perovskite structure continue to attract attention due to their ability to accommodate cation substitution and, consequently, modulate their magnetic and electronic characteristics. Compounds belonging to the families of $La_{1-x}Sr_xFeO_3$ and $La_{1-x}Sr_xCoO_3$ families have shown great potential for magnetocaloric applications due to their ability to exhibit ferromagnetic–paramagnetic transitions near room temperature and the possibility of tailoring their properties through cationic substitution [3,6–8].

The partial substitution of $Fe^{3+}$ by $Co^{3+}$ in $La_{0.6}Sr_{0.4}Fe_{1-x}Co_xO_3$ modifies the superexchange interactions ($Fe^{3+}$–O–$Fe^{3+}$, $Fe^{3+}$–O–$Co^{3+}$, $Co^{3+}$–O–$Co^{3+}$), leading to a progressive strengthening of the ferromagnetic order and a significant increase in $T_C$. At the same time, changes in ionic radii and spin states (low spin vs. high spin configurations) influence the local distortion of the $BO_6$ octahedra and the overall symmetry of the lattice.

Nanostructuring of these materials offers additional advantages. By confining crystallite growth using polymeric templates, it is possible to control particle size, morphology, and connectivity, which in turn affect magnetic anisotropy and the reversibility of the magnetocaloric cycle. Several studies have reported that nanostructured perovskites can exhibit enhanced MCE due to improved surface–interface exchange coupling and more efficient heat transfer [9–11].

In this work, we present a comprehensive structural, morphological, and magnetic characterization of the nanostructured series $La_{0.6}Sr_{0.4}Fe_{1-x}Co_xO_3$ (x = 0, 0.2, 0.5, 0.8 and 1), synthesized using polymeric membranes of 200 nm and 800 nm pore diameters as templates. The results reveal how both cobalt substitution and morphological control contribute to the magnetocaloric performance of these materials, providing valuable insight into the interplay between composition, structure, and magnetic entropy change.

## 2. Experimental

Nanostructured $La_{0.6}Sr_{0.4}Fe_{1-x}Co_xO_3$ (x = 0, 0.2, 0.5, 0.8 and 1) samples were prepared by the pore-wetting technique, following the methodology previously reported for $La_{0.6}Sr_{0.4}CoO_3$ perovskite tubes [12]. High-purity precursor salts $La(NO_3)_3{\cdot}6H_2O$, $Sr(NO_3)_2$, $Fe(NO_3)_3{\cdot}9H_2O$, and $Co(NO_3)_2{\cdot}6H_2O$ (Merck, 99.99%) were dissolved in distilled water to obtain 1 $mol{\cdot}L^{-1}$ stoichiometric solutions under acidic conditions (pH = 5).

Commercial polymeric membranes (Isopore™ by Millipore) with nominal pore diameters of 200 nm and 800 nm were used as templates. The precursor solution was introduced into the pores using a manual syringe-assisted infiltration system to ensure homogeneous filling. The filled membranes were partially dehydrated and denitrated using microwave heating, promoting the initial decomposition of nitrates and uniform distribution of precursors inside the pores.

The polymeric templates were then removed by calcination at 1000 °C for 2 h, yielding self-supported nanotubes or nanowires composed of perovskite nanoparticles forming the tube walls. The samples were labeled according to their nominal composition (x = 0, 0.2, 0.5, 0.8, and 1.0) and the pore diameter of the polymeric template (d = 200 nm and d = 800 nm).The synthesis route employed is a simple and effective method to obtain nanostructured perovskites with large surface areas and controlled morphology [13–15].

Qualitative and quantitative phase analyses were performed by X-ray powder diffraction (XPD). High-resolution synchrotron measurements were carried out at the D10B-XPD beamline of the Brazilian Synchrotron Light Laboratory (LNLS, Campinas, Brazil), using a high-intensity configuration with a wavelength of 1.7708 Å (incident energy ≈ 7000 eV). The diffraction patterns were analyzed by Rietveld refinement using the FullProf

Suite software [16], assuming the rhombohedral $R\bar{3}c$ structure (space group No. 167). In the refinement model, ($La^{3+}$, $Sr^{2+}$) cations occupy the 6a Wyckoff position, $Co^{3+}$/$Fe^{3+}$ the 6b site, and $O^{2-}$ anions the 18e position. The peak profiles were fitted using a pseudo-Voigt function, and the background was modeled by a six-parameter polynomial in $(2\theta)^n$ ($n$ = 0–5). Isotropic atomic displacement parameters were used, with equal values assigned to La and Sr atoms at the A-site. No additional constraints were imposed, and all refinements converged with excellent agreement factors. The average crystallite size was estimated via the Scherrer equation using the first intense low-angle reflection to minimize the influence of microstrain.

The morphology of the nanostructured powders was investigated by Scanning Electron Microscopy (SEM) using a Nova™ NanoSEM 230 (FEI), (Laboratorio de Microscopía Electrónica, Gerencia de Materiales, Centro Atómico Bariloche, CNEA, Argentina). The micrographs were obtained at accelerating voltages between 10 and 15 kV, and particle sizes were estimated by image analysis using ImageJ software. The images revealed hollow nanotube and nanowire structures with diameters consistent with the template pores (≈ 200–800 nm), and walls composed of interconnected nanoparticles within the 30–60 nm range.

Magnetic characterization was carried out using a VersaLab™ Vibrating Sample Magnetometer (Quantum Design) at (Laboratorio de Propiedades Eléctricas y Magnéticas de Óxidos Multifuncionales, Departamento de Física de la Materia Condensada at Centro Atómico Constituyentes, CNEA, Argentina). The magnetization (M) was measured under applied magnetic fields up to 3 T over the 50–400 K temperature range.

For each composition, Zero-Field-Cooled Warming (ZFCW), Field-Cooled Cooling (FCC), and Field-Cooled Warming (FCW) protocols were followed to evaluate magnetic reversibility and thermal stability. Isothermal M(H) curves were recorded at selected temperatures for both complete and first-quadrant field cycles, enabling the estimation of the magnetic entropy change ($-\Delta S_M$) and the Relative Cooling Power (RCP).

## 3. Results and discussion

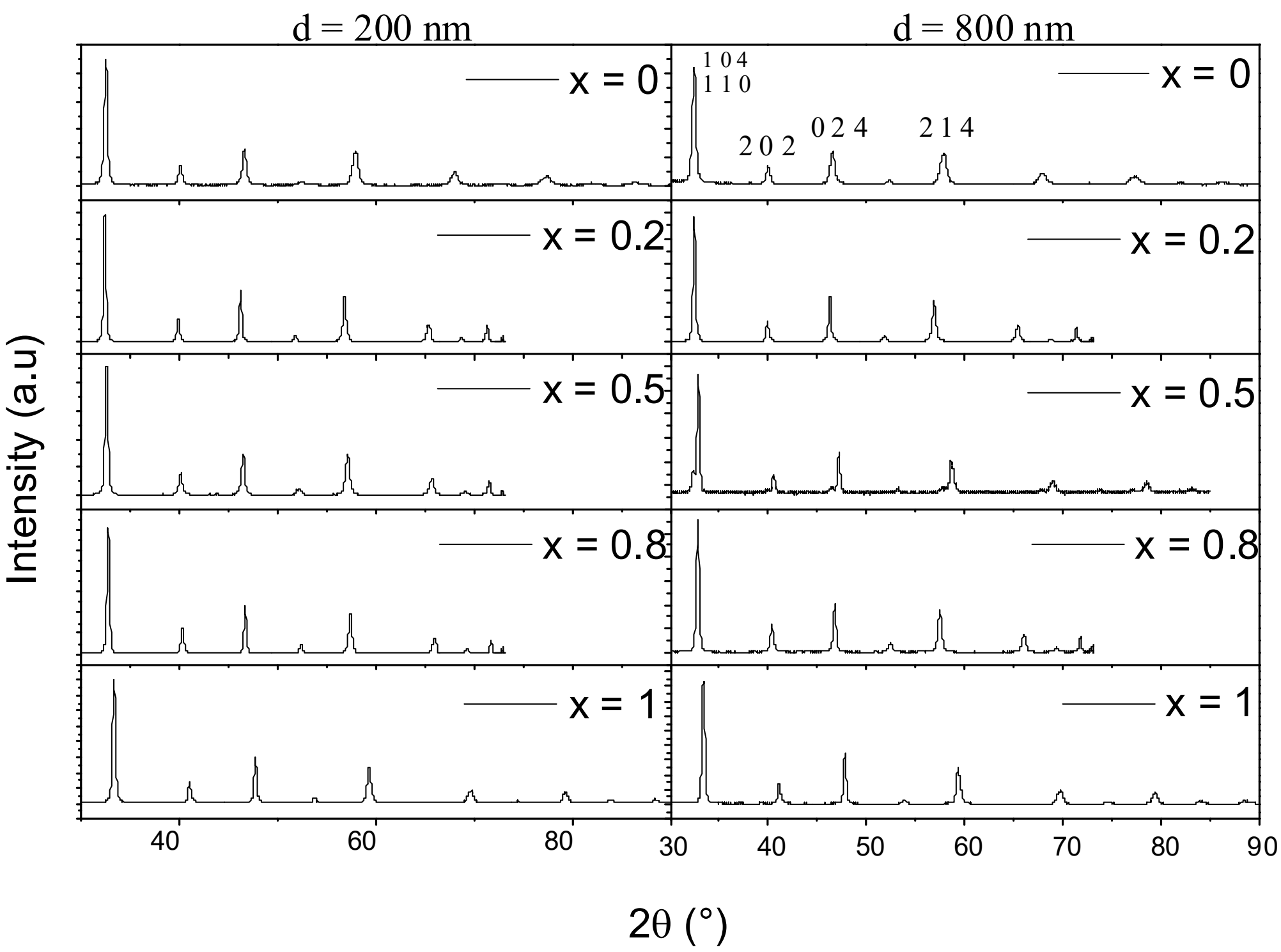

Fig. 1. X-ray diffraction patterns of all $La_{0.6}Sr_{0.4}Fe_{1-X}Co_xO_3$ samples.

Fig. 1 shows the X-ray diffraction (XRD) patterns obtained for the complete series of $La_{0.6}Sr_{0.4}Fe_{1-x}Co_xO_3$ samples (x = 0, 0.2, 0.5, 0.8, and 1.0), all calcined at 1000 °C and synthesized by the pore-wetting method using polymeric membranes with pore diameters of 200 nm and 800 nm.

All compositions exhibit diffraction peaks characteristic of a perovskite-type $ABO_3$ structure with distorted symmetry, corresponding to a single phase without detectable crystalline impurities. The main reflections are indexed to the corresponding planes. The most intense peaks are associated with the (110)/(104) planes (2θ ≈ 32.5–33°) and the (202)/(024) (2θ ≈ 47°), corresponding to a rhombohedral structure, with $R\bar{3}c$ space group.

Crystallite size values estimated using the Scherrer equation $d = \frac{k\lambda}{\beta \cos\theta}$ for the most intense reflections, are presented in table II. Values remain within the nanometric scale (30–60 nm), without indicating a clear trend with composition[17]. The preservation of this size scale, even after calcination at 1000 °C, is likely to be due to the confinement method, which limits the crystallite growth and promotes the stability of the nanostructures.

The scanning electron microscopy (SEM) micrographs shown in figure 2 reveal a systematic evolution of the morphology as a function of cobalt content (x) and thermal treatment.

For the undoped composition $x = 0$ with $d = 200$ nm, we observe a homogeneous distribution of thin wall nanotubes, with external diameters of approximately 167 nm and crystallite sizes around 39 nm. As the substitution of Fe by Co increases, an increment is observed in the size of the particles that form the nanostructures and the morphology evolves toward nanorods or nanowires, with the exception of $x=0.8$, for which an unclear morphology has been obtained. The morphological parameters, external diameters, and crystallite sizes for all samples are summarized in Table I.

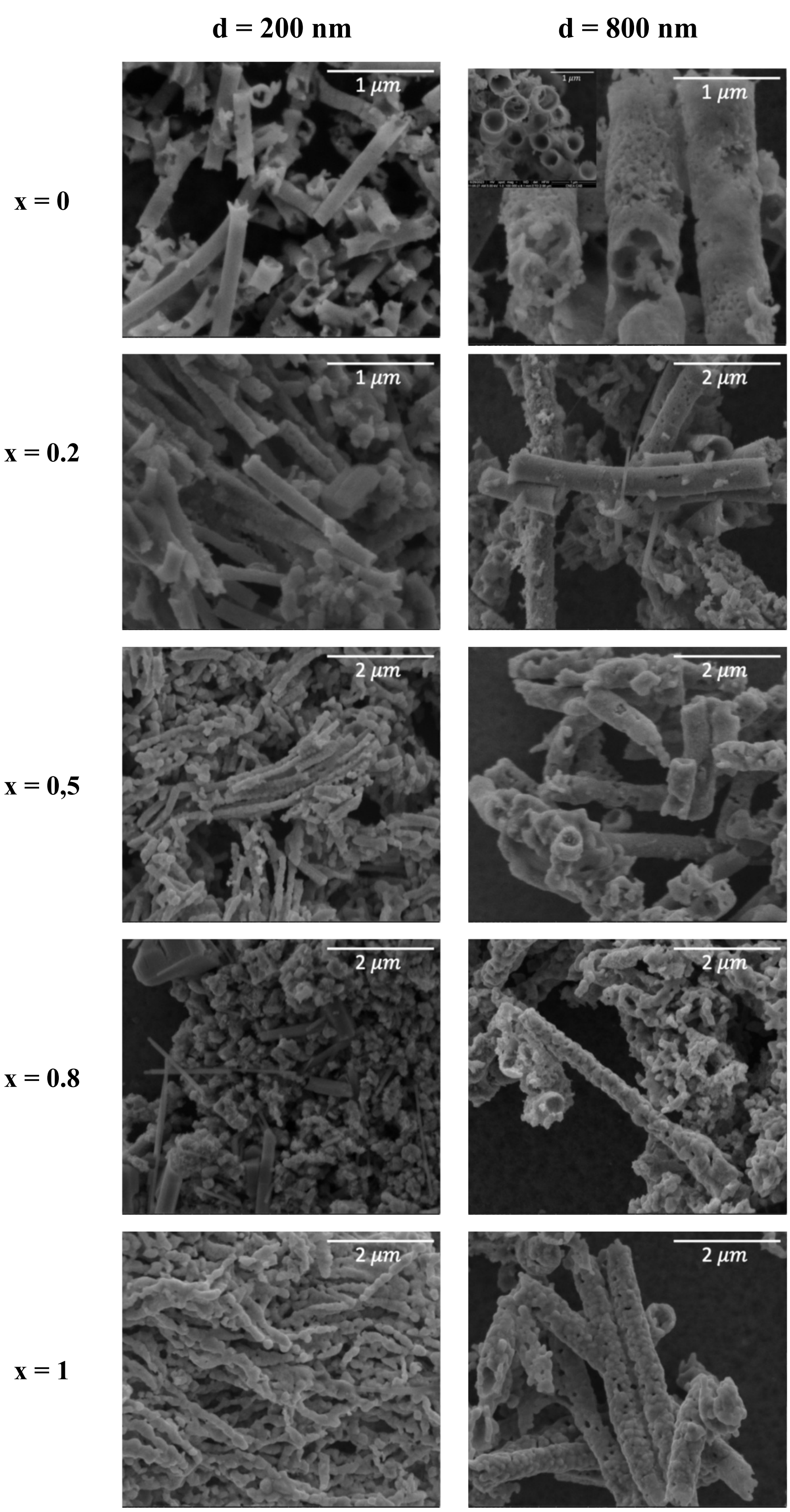

Figure 2. SEM micrographs of the $La_{0.6}Sr_{0.4}Fe_{1-x}Co_xO_3$ samples.

Samples synthesized using d = 800 nm pore membranes, exhibit nanotubes with hollow structures. For x = 0, 0.2 and 0.5 some internal cavities are observed inside the nanotubes. For x = 0.8 and 1, an increase in particle size and wall thickness is observed.

For both template sizes, the crystallite diameter seems almost unaffected.

**Table I. Morphological, particle size, and crystallite size data obtained from SEM micrographs and XRD patterns for the $La_{0.6}Sr_{0.4}Fe_{1-x}Co_xO_3$ samples.**

| d (nm) | x | Morphology | Outer Diameter (nm) | Crystallite Size (nm) |
|---|---|---|---|---|
| 200 | 0 | Nanotubes | 167 ± 2 | 38.8 ± 0.8 |
| | 0.2 | Nanotubes | 147 ± 3 | 61 ± 2 |
| | 0.5 | Nanowires | 156 ± 3 | 43.6 ± 0.9 |
| | 0.8 | unclear | -- | 58 ± 1 |
| | 1 | Nanowires | 135 ± 13 | 53 ± 2 |
| 800 | 0 | Nanotubes | 545 ± 2 | 30.9 ± 0.5 |
| | 0.2 | Nanotubes | 554 ± 6 | 55 ± 2 |
| | 0.5 | Nanotubes | 515 ± 2 | 59 ± 2 |
| | 0.8 | Nanotubes | 413 ± 8 | 54 ± 1 |
| | 1 | Nanotubes | 646 ± 48 | 56 ± 2 |

## 4. Magnetic and Magnetocaloric Properties

Figure 3 shows the magnetization versus temperature (M(T)) curves for the $La_{0.6}Sr_{0.4}Fe_{1-x}Co_xO_3$ series, measured under an applied magnetic field of 1000 Oe. A clear dependence of magnetization on cobalt content is observed. The samples with x = 1 exhibit the highest low-temperature magnetization (≈ 29–33 emu $g^{-1}$ at 10 K) and well-defined magnetic transitions near 240 K and 230 K, respectively. Intermediate compositions (x ≈ 0.5–0.8) display lower magnetization values at low temperature (12–20 emu $g^{-1}$) and broader transitions, indicative of increased magnetic inhomogeneity. Samples with low Co content (x ≤ 0.2) exhibit weak magnetic responses (< 1 emu $g^{-1}$) and no clear ferromagnetic transition within the measured range. A systematic difference is observed between the two series: samples with d = 800 nm tend to exhibit higher magnetization values than those d = 200 nm.

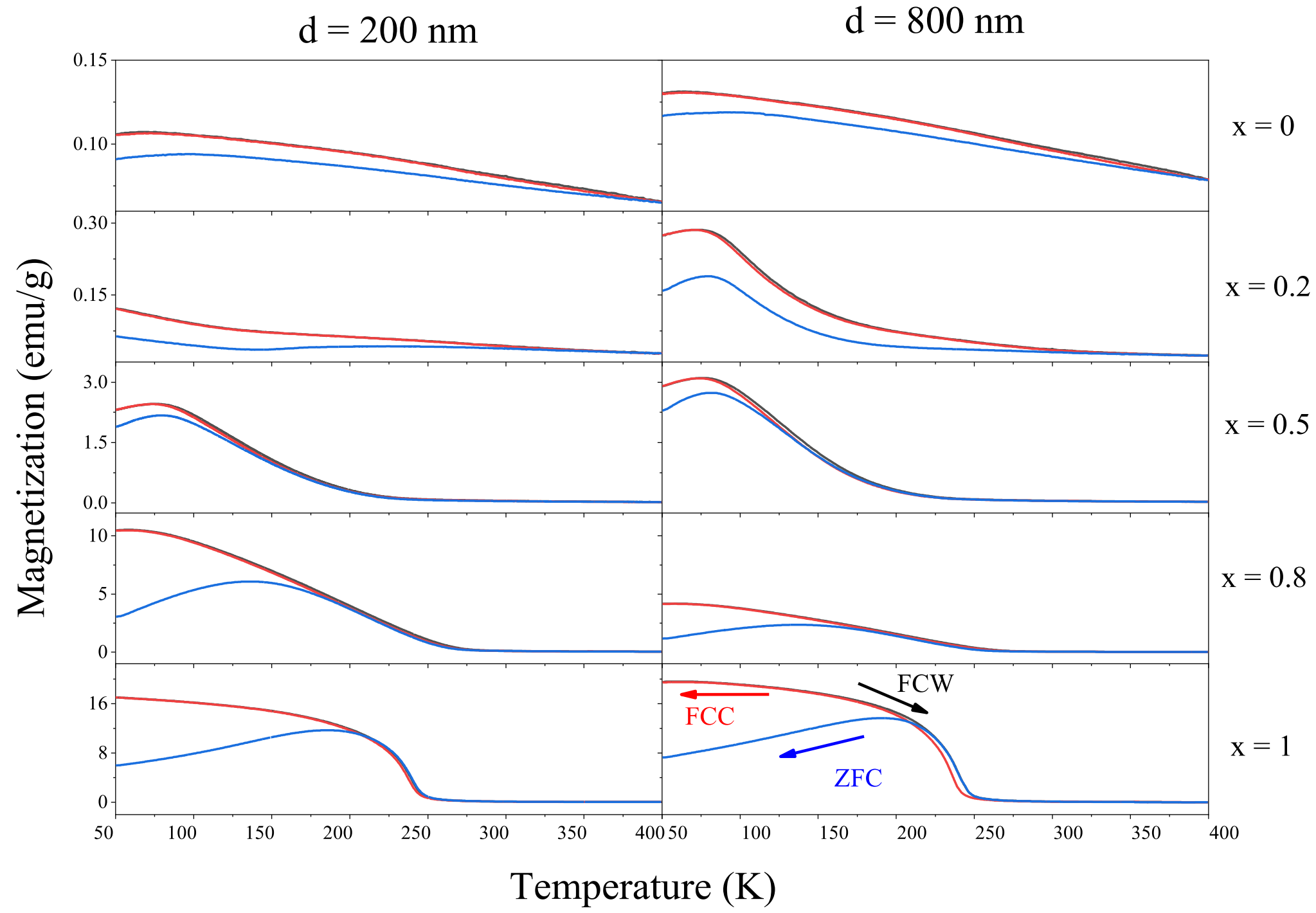

Figure 3. Magnetization versus temperature (M(T)) curves measured under an applied field of 1000 Oe for $La_{0.6}Sr_{0.4}Fe_{1-x}Co_xO_3$ samples.

Figs. 4 and 5 show the magnetization as a function of applied magnetic field (M(H)) at 50 K for the samples with d = 200 nm and d = 800 nm, respectively. A systematic increase in the maximum magnetization ($M_{max}$) is observed as the Co content increases, indicating that the gradual substitution of $Fe^{3+}$ by Co modifies the balance of magnetic exchange interactions and enhances the ferromagnetic component of the system. This trend can be rationalized considering the evolution of the dominant superexchange pathways: $Fe^{3+}$–O–$Fe^{3+}$ interactions are typically antiferromagnetic or magnetically weak, whereas $Fe^{3+}$–O–$Co^{3+}$ and $Co^{3+}$–O– $Co^{4+}$ linkages favor stronger ferromagnetic coupling. Similar competition between AFM ($Co^{3+}$–O–$Co^{3+}$, $Co^{3+}$–O–$Fe^{3+}$) and FM ($Co^{3+}$–O–$Co^{4+}$, $Fe^{3+}$–O–$Co^{4+}$) interactions has been reported in related La–Sr–Co–Fe perovskites [18], demonstrating that changes in the Co/Fe ratio strongly influence the emergence of ferromagnetic nanoregions. Moreover, as highlighted by Shinawi et al. [19], variations in the oxidation state and spin configuration of Co (particularly the stabilization of intermediate or high-spin $Co^{3+}$) can significantly strengthen ferromagnetic superexchange, providing further support for the observed increase in $M_{max}$ with Co content in the present series.

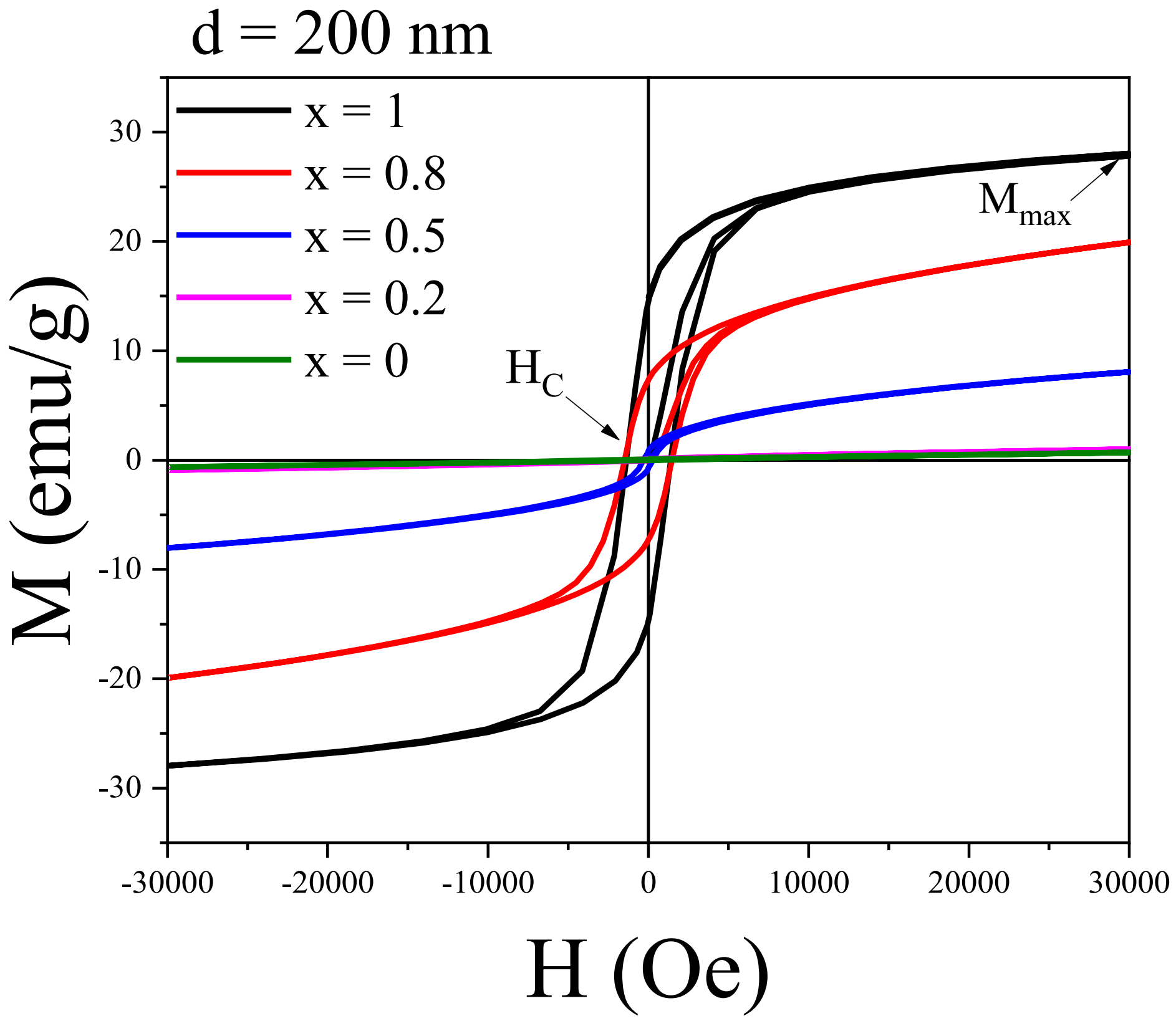

Figure 4. Magnetization versus magnetic field M(H) curves for samples with d = 200 nm, measured at 50 K.

In Fig. 4 with d = 200 nm, we observe that the maximum value of the magnetization increases monotonically from x = 0 to x = 1 displays the lowest.

The influence of morphology on the magnetic behavior must be interpreted with caution, since several microstructural and electronic factors evolve simultaneously during synthesis. Although samples with d = 800 nm generally exhibit higher magnetization than those obtained with d = 200 nm, this trend cannot be directly ascribed to geometrical features alone. Larger pore sizes typically lead to thicker nanotube walls and larger primary grains, which may improve magnetic exchange connectivity. However, grain coarsening is also accompanied by variations in cation distribution, local strain and oxygen vacancy concentration, all of which can substantially modify the delicate balance between ferromagnetic and antiferromagnetic interactions in Co-doped ferrites.

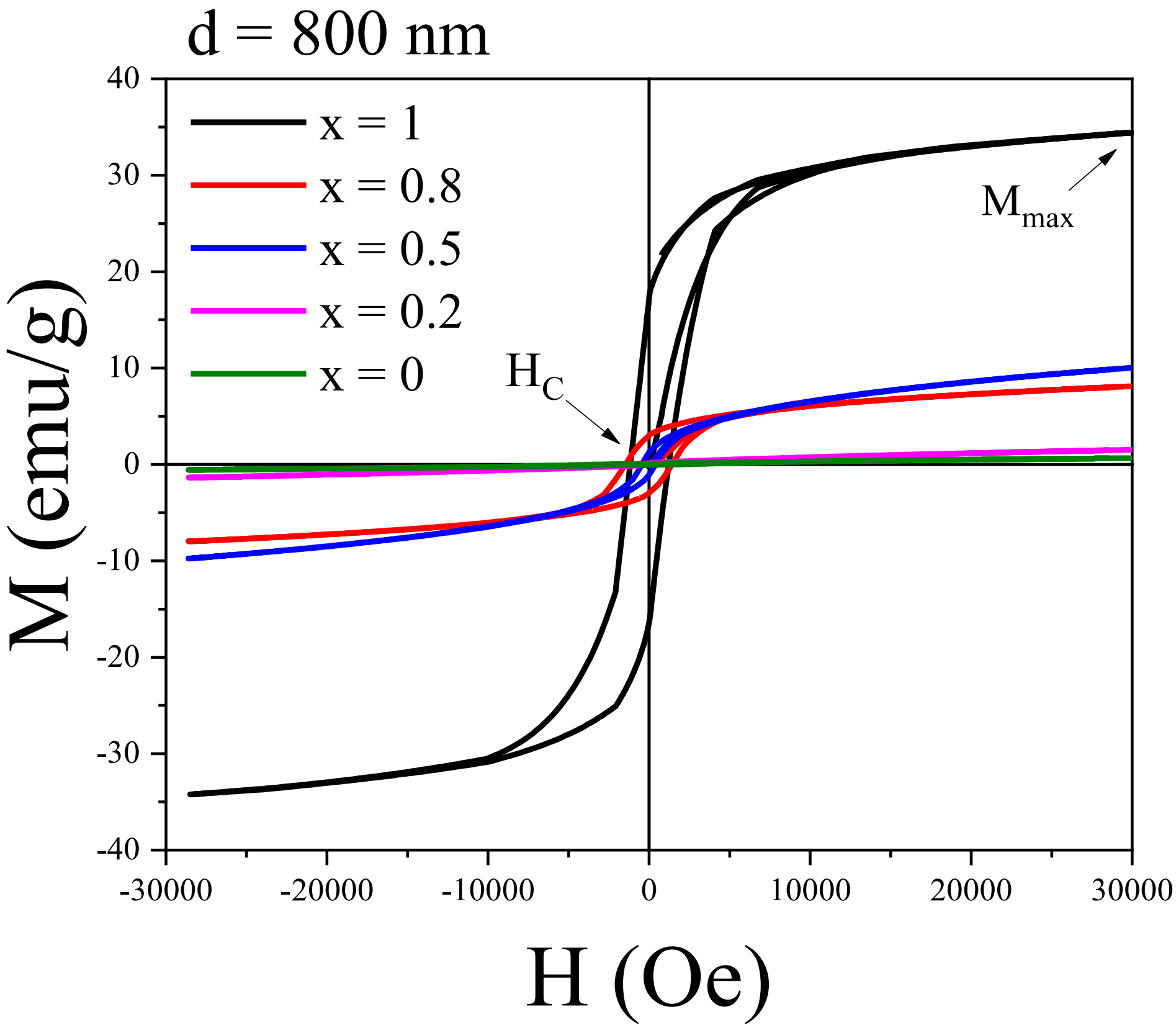

Figure 5. Magnetization versus magnetic field M(H) curves for samples with d = 800 nm, measured at 50 K.

Figure 5 reveals that, within d = 800 nm series, we also see an increase of the saturation of magnetization from x = 0 to x = 1. However, the evolution is less clear. For example, samples with x = 0.5 and x = 0.8, display similar M(H) values, both significantly lower than the one corresponding to x = 1, but higher than x = 0.

At low temperature, the coercive field ($H_C$) decreases systematically with increasing Co content (Fig. 5). The x = 0.5 sample exhibits the lowest $H_C$, which suggests a more homogeneous magnetic environment and fewer obstacles to a monodomain motion compared with Fe-rich compositions. In contrast, the x = 0.2 and x = 1 samples exhibit higher Hc and incomplete saturation. This behavior is consistent with oxygen non-stoichiometry, which modifies the $Fe^{3+}/Fe^{4+}$ and $Co^{3+}/Co^{4+}$ balance and generates locally frustrated magnetic regions. Oxygen vacancies disrupt superexchange pathways and promote spin-disordered clusters, acting as pinning centers that hinder full alignment under moderate fields. Consequently, the coexistence of FM and AFM interactions becomes more pronounced, increasing coercivity and limiting the approach to saturation [20].

To further investigate the magnetic transition type in the $La_{0.6}Sr_{0.4}Fe_{1-x}Co_xO_3$ series, we present Arrott plots ($M^2$ vs. H/M) at different temperatures near the magnetic transition, as shown in Fig. 6a) for samples using templates of d = 200 nm samples and Fig. 6b) for samples with d = 800 nm. By these plots, it is possible to distinguish between first- and second-order phase transitions following the Banerjee criterion, which states that positive slopes across all isotherms correspond to second-order transitions [21].

The Arrott plots for samples with d = 200 nm and d = 800 nm exhibit positive slopes throughout the measured field range, indicating second-order magnetic transitions according to the Banerjee criterion. However, the curvature and dispersion of the high-field region are pronounced for Fe-rich compositions, preventing a reliable determination of $T_C$ and revealing the presence of magnetic disorder. Only at higher Co content do some isotherms approach the expected linear behavior, although without a clear convergence toward the origin. This is consistent with the weak magnetization previously observed for Fe-rich samples.

In Table II, we present the slopes obtained from linear fitting of the high-field region, displaying a systematic variation with Co content. Samples with higher Co content (x = 1) exhibit smaller slopes (on the order of $10^{-8}$–$10^{-10}$ emu $g^{-1}$ $Oe^{-1}$), consistent with a more ideal mean-field-like behavior. In contrast, compositions with lower Co content (x = 0.2 – 0.5) show larger slopes ($10^{-2}$ – $10^{-3}$ emu $g^{-1}$ $Oe^{-1}$), indicative of increased magnetic disorder and less homogeneous ferromagnetic coupling.

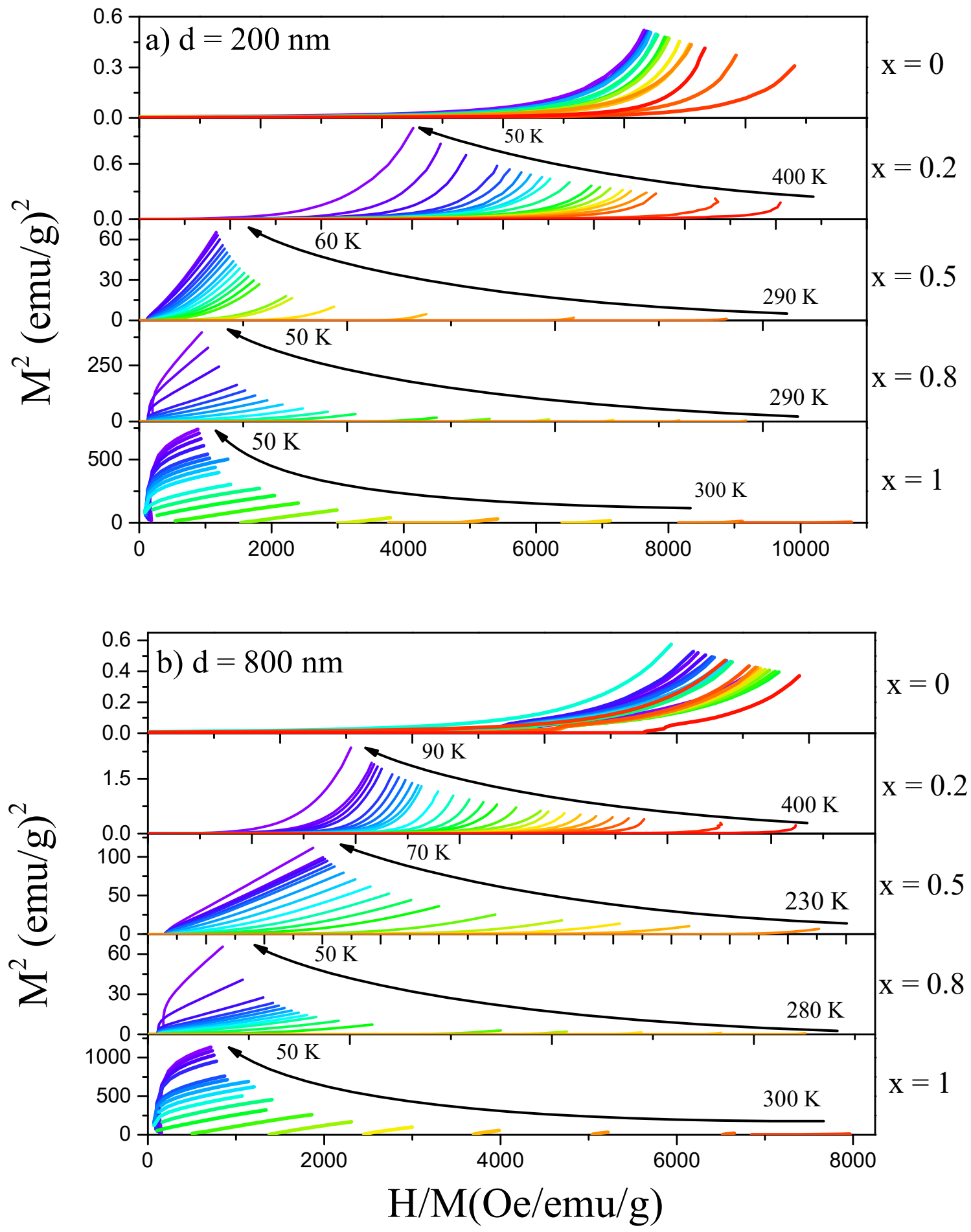

Figure 6. Arrott plots for $La_{0.6}Sr_{0.4}Fe_{1-x}Co_xO_3$ samples with a) d = 200 nm and b) d = 800 nm.

The Arrott plots of the d = 800 nm series (Fig. 6b)) display a more well-defined behavior at high magnetic fields, particularly for the fully Co-substituted compounds x = 1. In these samples, the extrapolated isotherms tend to intersect the $M^2$ axis at positive values closer to the origin, allowing a more confident estimation of $T_C$ in the range of ~230–240 K. Intermediate compositions (x ≈ 0.5–0.8) still show deviations from ideal linearity, but with noticeably reduced curvature compared to their d = 200 nm counterparts.

Overall, these observations indicate that Co substitution is the primary factor driving the stabilization of long-range ferromagnetic order, while the improved definition of the Arrott plots in the d = 800 nm series suggests that morphology may assist by providing better magnetic connectivity. Nevertheless, additional structural and spectroscopic characterization is required to disentangle the specific contributions of pore-driven microstructural changes from those directly associated with Co incorporation.

**Table II. Data obtained from Arrott plot analysis.**

| d (nm) | x | $T_C$ (K) | Slope (emu $g^{-1}$ $Oe^{-1}$) |
|---|---|---|---|
| 200 | 0 | — | — |
| | 0.2 | — | $1.52 \times 10^{-8}$ |
| | 0.5 | 80 | $1.41 \times 10^{-8}$ |
| | 0.8 | 150 | $6.54 \times 10^{-8}$ |
| | 1 | 240 | $3.74 \times 10^{-8}$ |
| 800 | 0 | — | — |
| | 0.2 | — | $3.18 \times 10^{-8}$ |
| | 0.5 | 70 | $3.96 \times 10^{-8}$ |
| | 0.8 | 165 | $3.41 \times 10^{-8}$ |
| | 1 | 230 | $2.15 \times 10^{-8}$ |

Figure 7 show the magnetic entropy change ($-\Delta S_M$) as a function of temperature for the $La_{0.6}Sr_{0.4}Fe_{1-x}Co_xO_3$ series under an applied field of 3 T. The $\Delta S_m$ values were derived from the M(H) curves using the Maxwell relation, revealing the influence of cobalt doping and the morphology, on the magnetocaloric effect (MCE).

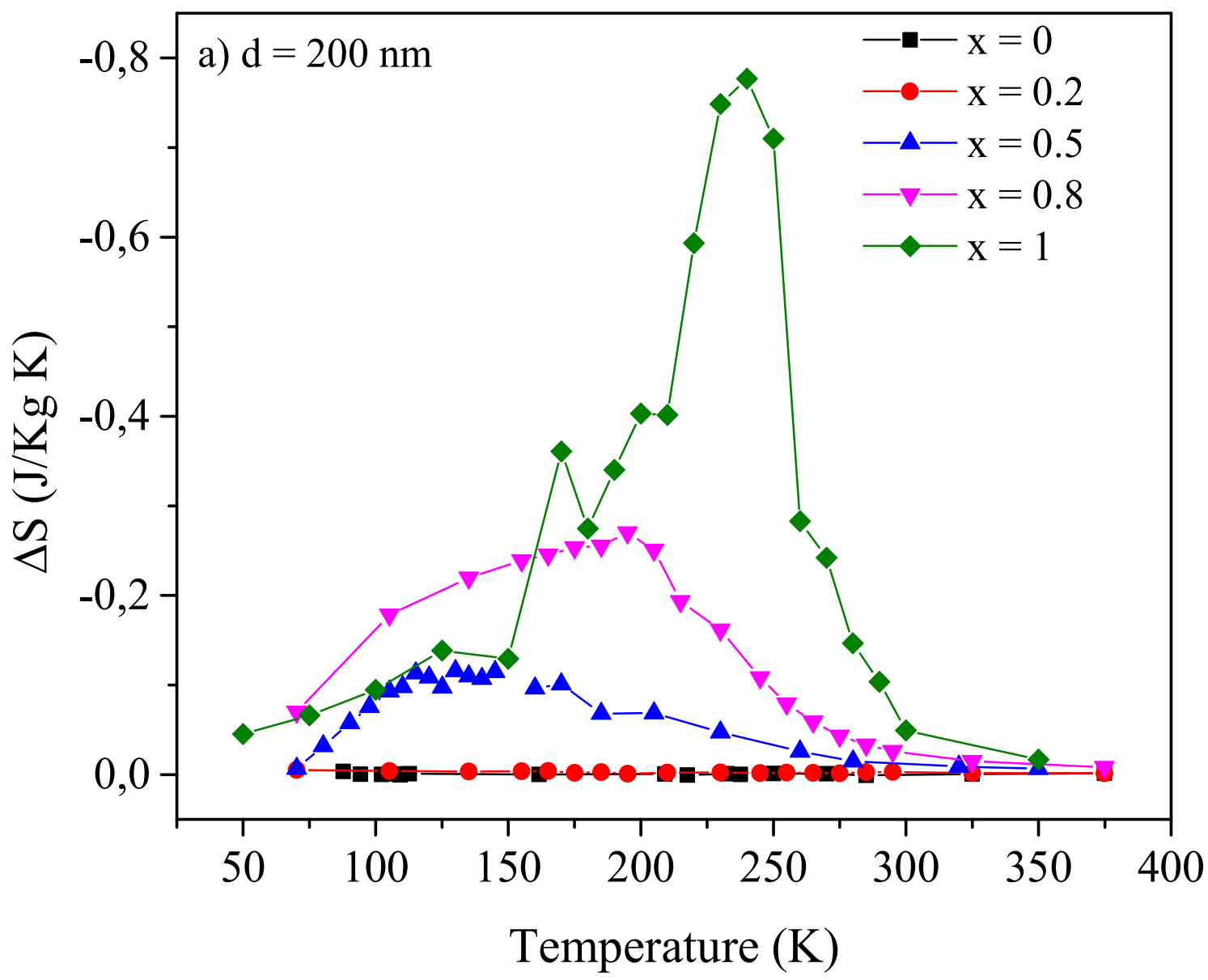

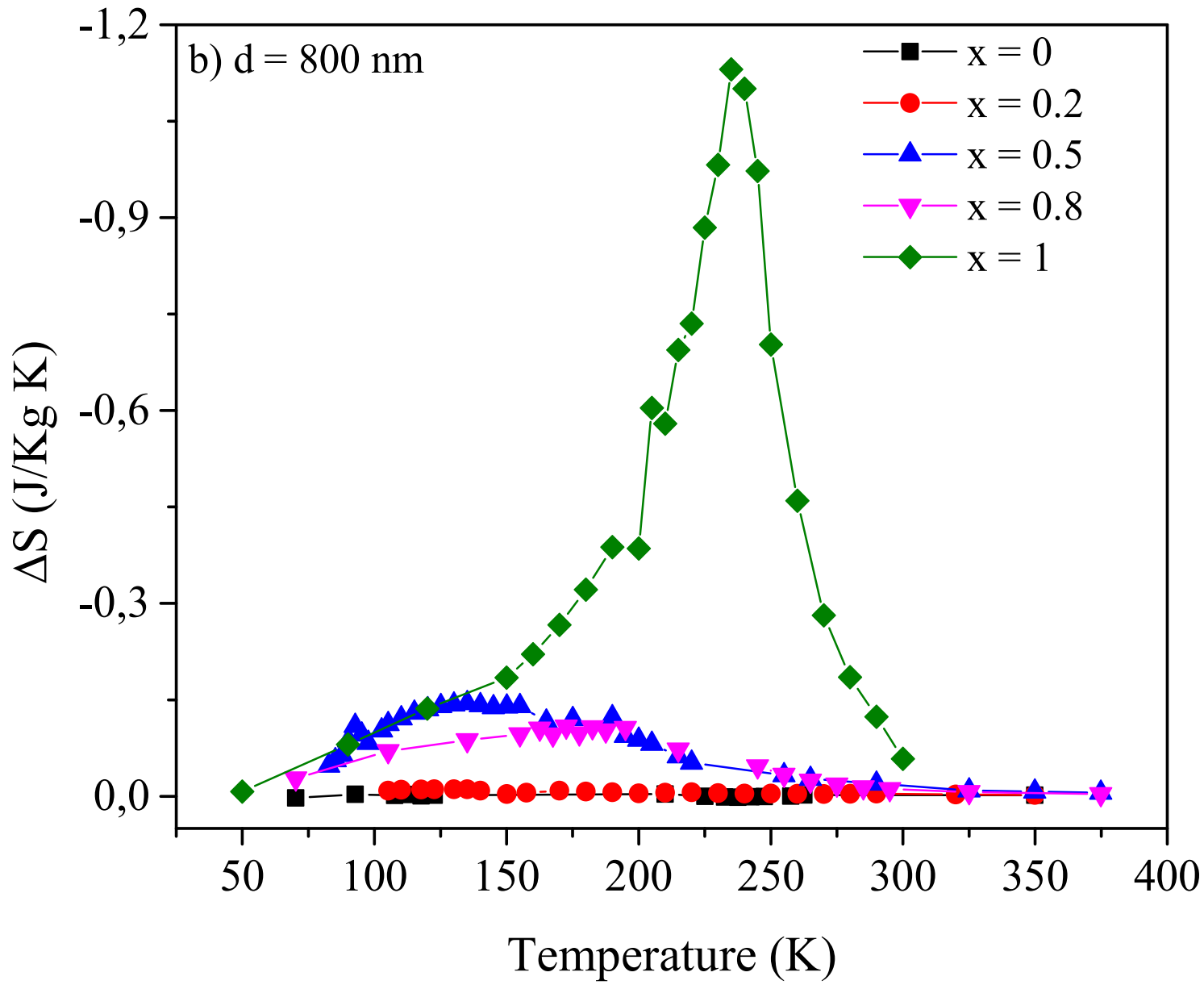

Figure 7. Magnetic entropy change ($-\Delta S_M$) as a function of temperature for a) the d = 200 nm series and b) the d = 800 nm series.

**Table III. MCE parameters for all samples at H = 30 kOe.**

| d (nm) | x | $-\Delta S_{max}$ (J kg$^{-1}$ K$^{-1}$ × $10^{-3}$) | $T_{peak}$ (K ± 5) | RCP (J kg$^{-1}$) |
|---|---|---|---|---|
| 200 | 0 | Negligible | Negligible | Negligible |
| | 0.2 | 0.002 | 185 | Negligible |
| | 0.5 | 0.116 | 130 | 13.3 ± 0.5 |
| | 0.8 | 0.270 | 195 | 37 ± 1 |
| | 1 | 0.770 | 240 | 35 ± 4 |
| 800 | 0 | Negligible | Negligible | Negligible |
| | 0.2 | 0.010 | 165 | Negligible |
| | 0.5 | 0.107 | 135 | 11.7 ± 0.4 |
| | 0.8 | 0.145 | 183 | 15.8 ± 0.6 |
| | 1 | 1.130 | 240 | 59 ± 5 |

For the d = 200 nm series, a clear increase in $-\Delta S_{max}$ is observed with increasing Co content, in the sense in which the largest peak corresponds to sample x = 1, reaching a value of $-\Delta S_{max}$ = 0.770 J kg$^{-1}$ K$^{-1}$ at 240 K (correlated to the ferromagnetic transition). As Co content decreases, the magnitude of the effect diminishes. Sample x = 0.8 exhibits $-\Delta S_{max}$ = 0.270 J kg$^{-1}$ K$^{-1}$ at 195 K, x = 0.5 reaches $-\Delta S_{max}$ = 0.116 J kg$^{-1}$ K$^{-1}$ at 130 K while x = 0.2 sample presents an almost negligible signal ($-\Delta S_{max}$ = 0.002 J kg$^{-1}$ K$^{-1}$). Consistently, this last sample presents a weak magnetic response.

A similar trend is observed in the d = 800 nm series, though with higher overall magnetic entropy values. The x = 1 sample exhibits the largest MCE in the entire series, reaching a peak of $-\Delta S_{max}$ = 1.130 J kg$^{-1}$ K$^{-1}$ centered at 240 K. For x = 0.8, the peak decreases to 0.145 J kg$^{-1}$ K$^{-1}$ at 183 K, while x = 0.5 shows $-\Delta S_{max}$ = 0.107 J kg$^{-1}$ K$^{-1}$ at 135 K. The lightly doped sample with x = 0.2 exhibits a small and broad entropy change of 0.010 J kg$^{-1}$ K$^{-1}$.

Comparison between the two morphological groups reveals that, for a given composition, d = 800 nm series display both higher $-\Delta S_{max}$ values and broader transition profiles, for the highest Co contents. This behavior suggests that in theses samples, an open interconnected morphology enhances magnetic connectivity and increases the effective volume participating in the field-induced entropy change, leading to improved MCE efficiency.

The refrigerant capacity, or relative cooling power (RCP), was also evaluated, as summarized in Table III. It is defined as the product of $-\Delta S_{max}$ and the full width at half maximum ($\delta T_{FWHM}$) of the entropy change peak. The d = 800 nm with x = 1 sample stands out with an RCP of 59 J $kg^{-1}$. The RCP diminishes while increasing iron doping. This trend is not that clear in the samples obtained from templates of d = 200 nm, in which samples with x = 1 and 0.8 display similar RCP values. However, a clear reduction of RCP is observed for x = 0.5.

These results points to the fully Co-doped compositions as the most efficient in terms of refrigerant capability, with transitions centered around 230–240 K. Intermediate compositions x = 0.5 show lower RCP values (13.3 and 11.7 J $kg^{-1}$, respectively), whereas samples with x = 0.2 show negligible MCE values.

Taken together, these findings demonstrate that increasing Co content not only enhances the magnitude of the magnetocaloric effect but also increases its effective temperature range, thereby improving the overall cooling efficiency of the material.

## 5. Conclusions

A comprehensive study of the $La_{0.6}Sr_{0.4}Fe_{1-x}Co_xO_3$ (x = 0, 0.2, 0.5, 0.8, and 1.0) nanostructured perovskite series synthesized by the pore-wetting method has been presented. Structural, morphological, and magnetic analyses demonstrate the close relationship between composition, morphology, and magnetocaloric performance.

X-ray diffraction confirmed the formation of a single-phase perovskite structure with rhombohedral symmetry (space group $R\bar{3}c$) for all compositions, without detectable secondary phases. The average crystallite size remained within the nanometric range (30–60 nm) across the series, even after calcination at 1000 °C, highlighting the efficiency of morphological confinement in controlling crystal growth.

Morphological characterization revealed a progressive evolution from thin nanotubes in Fe-rich compositions to thicker, more porous nanowires and nanotubes with increasing Co content. The use of polymeric membranes with different pore sizes (200 and 800 nm) allowed control over particle allowing the preservation of nanostructure, together with the presence of interconnected pores and multiple internal cavities in several samples, enhances magnetic connectivity and facilitates heat and field driven diffusion features that are particularly advantageous for magnetocaloric applications.

Magnetic measurements showed that cobalt substitution strengthens ferromagnetic coupling, resulting in increased saturation magnetization ($M_S$) and higher Curie temperatures ($T_C$). The magnetic transitions were identified as second order according to the Banerjee criterion, consistent with the smooth M(T) behavior and reversible magnetization curves.

The magnetocaloric results revealed that Co content plays a decisive role in enhancing the magnetic entropy change ($-\Delta S_M$) and relative cooling power (RCP). The C8T2 sample (x = 1, 800 nm membrane pore size) exhibited the best overall performance, with $-\Delta S_{max}$ = 1.13 J kg$^{-1}$ K$^{-1}$ and RCP = 59 J kg$^{-1}$ under 3 T, indicating efficient and reversible magnetic refrigeration behavior near 240 K. Nanostructuring, undoubtedly plays a role related to the feasibility to control the surface/volume ratio, and thus, the connectivity with the potential reservoir to be thermalized

These findings confirm that combining controlled Co doping with nanoscale morphological design represents an effective strategy to optimize the magnetocaloric effect in perovskite-type oxides. The results also provide valuable insights into the relationship between structural distortion, magnetic order, and thermomagnetic response in transition-metal-based perovskites.

The combined effect of Co substitution and nanostructure control allows the understanding on how to tune both magnetic and magnetocaloric properties, an extremely important aspect regarding intermediate-temperature magnetic refrigeration.

**Acknowledgements**

The authors acknowledge financial support from Agencia Nacional de Promoción Científica y Tecnológica (Argentina) under projects ANPCyT (PICT 2021–00495). The authors also thank Gustavo Segovia for his support in graphic design preparation.

**CRediT authorship contribution statement**

**F. Morales:** Writing – review & editing, Writing – original draft, Methodology, Investigation, Formal analysis. **M. Quintero:** Writing – review & editing, Resources,

Methodology, Investigation, Formal analysis. **J. Sacanell:** Writing – review & editing, Resources, Methodology, Investigation, Formal analysis.

## 6. Bibliography

[1] A. Greco, C. Aprea, A. Maiorino, C. Masselli, A review of the state of the art of solid-state caloric cooling processes at room-temperature before 2019, Int. J. Refrig. 106 (2019) 66–88. https://doi.org/10.1016/j.ijrefrig.2019.06.034.

[2] P.Z.Z. Nehan, O. Vitayaya, D.R. Munazat, M.T.E. Manawan, D. Darminto, B. Kurniawan, The magnetocaloric effect properties for potential applications of magnetic refrigerator technology: a review, Phys. Chem. Chem. Phys. 26 (2024) 14476–14504. https://doi.org/10.1039/D4CP01077A.

[3] H. Biswal, T.R. Senapati, A. Haque, J.R. Sahu, Beneficial effect of Mn-substitution on magnetic and magnetocaloric properties of La0.5Sr0.5CoO3 ceramics, Ceram. Int. 46 (2020) 11828–11834. https://doi.org/10.1016/j.ceramint.2020.01.217.

[4] P.T. Long, T. V. Manh, T.A. Ho, V. Dongquoc, P. Zhang, S.C. Yu, Magnetocaloric effect in La1-xSrxCoO3 undergoing a second-order phase transition, Ceram. Int. 44 (2018) 15542–15549. https://doi.org/10.1016/j.ceramint.2018.05.216.

[5] M. Imada, A. Fujimori, Y. Tokura, Metal-insulator transitions, 70 (1998) 1039–1263.

[6] P.T. Long, T. V. Manh, T.A. Ho, V. Dongquoc, P. Zhang, S.C. Yu, Magnetocaloric effect in La1-xSrxCoO3 undergoing a second-order phase transition, Ceram. Int. 44 (2018) 15542–15549. https://doi.org/10.1016/j.ceramint.2018.05.216.

[7] M. Itoh, I. Natori, S. Kubota, K. Motoya, Spin-Glass Behavior and Magnetic Phase Diagram of La1-xSrxCoO3 ( 0 ≤x ≤0.5) Studied by Magnetization Measurements, J. Phys. Soc. Japan. 63 (1994) 1486–1493. https://doi.org/10.1143/JPSJ.63.1486.

[8] L. Joshi, S. Keshri, Magneto-transport properties of Fe-doped LSMO manganites, (n.d.). https://doi.org/10.1016/j.measurement.2011.02.005.

[9] A. Kumar, D. Sivaprahsam, A.D. Thakur, Improvement of thermoelectric properties of lanthanum cobaltate by Sr and Mn co-substitution, J. Alloys Compd. 735 (2018) 1787–1791. https://doi.org/10.1016/J.JALLCOM.2017.11.334.

[10] N. Raghu Ram, M. Prakash, U. Naresh, N. Suresh Kumar, T. Sofi Sarmash, T. Subbarao, R. Jeevan Kumar, G. Ranjith Kumar, K. Chandra Babu Naidu, Review on Magnetocaloric Effect and Materials, J. Supercond. Nov. Magn. 31 (2018) 1971–1979. https://doi.org/10.1007/s10948-018-4666-z.

[11] W. Zhong, C.T. Au, Y.W. Du, Review of magnetocaloric effect in perovskite-type oxides, Chinese Phys. B. 22 (2013). https://doi.org/10.1088/1674-1056/22/5/057501.

[12] F.M. Alvarez, M.B. Vigna, M. Quintero, D.G. Lamas, J. Sacanell, Magnetocaloric effect of nanostructured La0.6Sr0.4CoO3, J. Alloys Compd. 970 (2024). https://doi.org/10.1016/j.jallcom.2023.172507.

[13] J. Sacanell, M.G. Bellino, D.G. Lamas, A.G. Leyva, Synthesis and characterization La0.6Sr0.4CoO3 and La0.6Sr0.4Co0.2Fe0.8O 3 nanotubes for cathode of solid-oxide fuel cells, Phys. B Condens. Matter. 398 (2007) 341–343. https://doi.org/10.1016/j.physb.2007.04.039.

[14] A. Mejía Gómez, J. Sacanell, A.G. Leyva, D.G. Lamas, Performance of La0.6Sr0.4Co1−yFeyO3 (y=0.2, 0.5 and 0.8) nanostructured cathodes for intermediate-temperature solid-oxide fuel cells: Influence of microstructure and composition, Ceram. Int. 42 (2015) 3145–3153. https://doi.org/10.1016/j.ceramint.2015.10.104.

[15] J. Sacanell, A.G. Leyva, M.G. Bellino, D.G. Lamas, Nanotubes of rare earth cobalt oxides for cathodes of intermediate-temperature solid oxide fuel cells, J. Power Sources. 195 (2010) 1786–1792. https://doi.org/10.1016/j.jpowsour.2009.10.049.

[16] J. Rodríguez-Carvajal, Recent advances in magnetic structure determination by neutron powder diffraction, Phys. B Phys. Condens. Matter. 192 (1993) 55–69. https://doi.org/10.1016/0921-4526(93)90108-I.

[17] A.S. Khan, M.F. Nasir, M.T. Khan, A. Murtaza, M.A. Hamayun, Study of structural, magnetic and radio frequency heating aptitudes of pure and (Fe-III)

doped manganite (La1-x SrxMnO3) and their incorporation with Sodium Poly-Styrene Sulfonate (PSS) for magnetic hyperthermia applications, Phys. B Condens. Matter. 600 (2021) 412627. https://doi.org/10.1016/j.physb.2020.412627.

[18] V.S. Pokatilov, V.S. Rusakov, A.O. Makarova, V. V. Pokatilov, M.E. Matsnev, Specific features of magnetic states of impurity iron ions in the perovskite La0.75Sr0.25Co0.9857Fe0.02O3, Phys. Solid State. 58 (2016) 315–318. https://doi.org/10.1134/S1063783416020220.

[19] H. El Shinawi, J.F. Marco, F.J. Berry, C. Greaves, LaSrCoFeO5, LaSrCoFeO5F and LaSrCoFeO5.5: New La-Sr-Co-Fe perovskites, J. Mater. Chem. 20 (2010) 3253–3259. https://doi.org/10.1039/b927141d.

[20] M.A. Señarís-Rodríguez, J.B. Goodenough, Magnetic and Transport Properties of the System La1-xSrxCoO3-δ (0 < x ≤ 0.50), J. Solid State Chem. 118 (1995) 323–336. https://doi.org/10.1006/jssc.1995.1351.

[21] B.K. Banerjee, On a generalised approach to first and second order magnetic transitions, Phys. Lett. 12 (1964) 16–17. https://doi.org/10.1016/0031-9163(64)91158-8.